\documentclass[preprint]{aastex}

\slugcomment{MSUPHY00.02}

\begin{document}

\title{Low Frequency Gravitational Waves from White Dwarf MACHO
Binaries}

\author{William A. Hiscock, Shane L. Larson, Joshua R. Routzahn}
\affil{Department of Physics, Montana State University, Bozeman,
Montana 59717}

\and
\author{Ben Kulick}

\affil{Department of Physics, California Institute of Technology,
Pasadena, California 91125}



\begin{abstract}
The possibility that Galactic halo MACHOs are white dwarfs has
recently attracted much attention.  Using the known properties of
white dwarf binaries in the Galactic disk as a model, we estimate
the possible contribution of halo white dwarf binaries to the
low-frequency ($10^{-5}\; {\rm Hz}\, <\, f\, <\, 10^{-1}\; {\rm
Hz}$) gravitational wave background.  Assuming the fraction of
white dwarfs in binaries is the same in the halo as in the disk,
we find the confusion background from halo white dwarf binaries
could be five times stronger than the expected contribution from
Galactic disk binaries, dominating the response of the proposed
space based interferometer LISA. Low-frequency gravitational wave
observations will be the key to discovering the nature of the
dark MACHO binary population.
\end{abstract}


\section{Introduction}\label{sec:Intro}
The MACHO project has detected 13-17 gravitational microlensing events
in the direction of the Large Magellanic Cloud to date (Alcock {\it et
al.}, 2000), while the EROS collaboration has detected two (Lasserre
{\it et al.}, 1999).  One result of the analysis of the observations
(Alcock {\it et al.}, 2000) is that, independent of the halo model
assumed, there are of order $2 \times 10^{11}$ MACHOs of mean mass
$0.5 M_\odot$ in the Galactic halo.\footnote{It is noteworthy that
this number is similar in magnitude to the number of stars in the
Galaxy.}

While the microlensing observations have confirmed the existence of
MACHOs (Massive Astrophysical Compact Halo Objects) in the halo, the
actual nature of these objects is still a matter of much debate.  The
mass suggests main sequence red dwarf stars, but these appear to be
ruled out observationally (Bahcall {\it et al.}, 1994; Graff \&
Freese, 1996a; Graff \& Freese, 1996b).  Until recently, white dwarf
stars also seemed to be a highly unlikely MACHO candidate, since
observations of the Galactic halo (see, {\it e.g.}, Flynn {\it et
al.}, 1996) did not find a population of old, red, white dwarf stars.
Since conventional baryonic stars seemed to be ruled out, highly
speculative candidate objects have been proposed as MACHO models, such
as primordial black holes and boson stars.

Recently, however, new models of white dwarf cooling processes
indicate that old, cool, white dwarf atmospheres form molecular
hydrogen, which can strongly absorb red light.  This implies that
aging white dwarfs will be blue objects, rather than red as was
previously believed (Saumon \& Jacobson, 1999; Hansen 1999).  In light
of these new cooling models, predictions have been made for the number
of halo white dwarfs that should be seen in deep field surveys
(Richer, 1999), and new analyses have detected candidate halo white
dwarfs with blue colors in the Hubble Deep Field (Ibata {\it et al.},
1999; M\'endez \& Minniti, 2000).  In addition, local blue white
dwarfs with large proper motions, indicating that they are members of
the halo population, have been tentatively identified (Ibata {\it et
al.}, 2000).

In this Letter we assume that the halo MACHOs are white dwarf
(WD) stars, and estimate the low frequency ($10^{-5}\; {\rm Hz} <
f < 10^{-1}\; {\rm Hz}$) gravitational wave (GW) background that
would be produced from a halo population of white dwarf binaries.
Such a background could be an important source for space-based
laser interferometer gravitational wave detectors such as the
proposed LISA mission (Bender {\it et al.}, 1998).  Gravitational
waves from halo WD binaries could be a very interesting signal
(if one is interested in characterizing the binary population of
MACHOs) or a serious confusion noise source (if one is concerned
that this background is masking signals from other sources of
interest, such as a cosmological background of GW). Estimates of
the gravitational wave signal from a halo composed of primordial
black holes has shown that the signal from binaries in the halo
could dominate the output of an interferometer such as LISA
(Hiscock, 1998; Ioka {\it et al.}, 1999).

In the absence of any observational evidence concerning the properties
a halo population of binary WDs, we make the assumption that white
dwarf binary properties in the halo mimic those of Galactic disk WD
binaries (Hils, Bender \& Webbink, 1990; Bender \& Hils, 1997).  This
assumption is almost certainly incorrect; the halo WD population is
generally felt to be much older than the disk population, and (based
on the microlensing events) probably has a mass distribution that
differs from the disk WD population.  However, using the disk as a
model is the best one can currently do.

\section{Disk binaries as GW sources}
The GW background generated by disk binaries (both Galactic and
extragalactic) has been thoroughly studied by Hils, Bender, and
Webbink (Hils, Bender \& Webbink, 1990; Bender \& Hils, 1997).
They have made careful estimates of the GW signal due to Galactic
disk WD binaries and also that due to extragalactic binaries. In
recent work, they have combined these signal amplitudes with the
planned properties of LISA to generate a simulation of LISA's
response to the combined Galactic and extragalactic signals.  The
key factor in this analysis is the width of a frequency bin in
the LISA data analysis for periodic sources. With a one-year
integration time, each frequency bin will have a width $\Delta f
= (1 \; {\rm yr})^{-1} \simeq 3 \times 10^{-8}\; {\rm Hz}$.  At
frequencies beginning at around $1 \times 10^{-3}$ Hz and higher,
the number of Galactic binaries per bin will begin to drop to
order unity. At this point, the properties of individual Galactic
binaries can be determined and their signal removed from the
record, so that the weaker combined signal of extragalactic
binaries will begin to be observable in the open bins.  Solving
for a particular Galactic binary and removing it from the data
record will typically require three bins of information
(Hellings, 1996). The effective spectral amplitude observed by
LISA, $h_f$, after taking the finite bin width into account, is
given by (Bender \& Hils, 1997)
\begin{equation}
    h_f = \left ( {(h_f^e)^2 \left [(h_f^G)^2 + (h_f^e)^2
    (1-e^{-3r})\right] \over
    e^{-3r}(h_f^G)^2+(h_f^e)^2(1-e^{-3r})}\right )^{1/2} \; \; ,
    \label{effhf}
\end{equation}
where $r$ is the number of Galactic binaries per frequency bin,
$h_f^G$ is the spectral amplitude of the Galactic binary background,
and $h_f^e$ is the spectral amplitude of the extragalactic binaries.

Using the relation between the spectral amplitude and the number
of binaries per unit frequency $dN/df$, the GW luminosity of a
binary $L(f)$, and the average inverse distance squared $\langle
d^{-2}\rangle$,
\begin{equation}
    h_f(f) = {2 \over \pi f} \left [ \left\langle {1 \over d^2}
    \right\rangle L(f) {dN \over df} \right ]^{1/2} \; \; ,
  \label{specamp}
\end{equation}
it is possible to extract an estimate of $dN/df$ from the
Bender-Hils results, along with an approximation to the spectral
amplitudes of the backgrounds due to Galactic disk binaries and
extragalactic binaries.  We find that the Bender-Hils results are
well approximated by
\begin{equation}
   {dN \over df} \simeq 4.47 \times 10^{-3} f^{-11/3} \; \; ,
   \label{dNdf}
\end{equation}
\begin{equation}
   \log_{10} \left (h_f^G \right ) = - \left ( {7 \over 6} \right )
   \log_{10} (h_f) - 21.8 \; \; ,
   \label {hfGdisk}
\end{equation}
and
\begin{equation}
  \log_{10} \left (h_f^e \right ) = - \left ( { 7 \over 6 } \right )
  \log_{10} (h_f) - 23.0 \; \;
  \label{hfedisk}
\end{equation}
where the frequency $f$ is measured in Hz. The frequency
dependence in these equations is characteristic of a population
of circular binaries that is evolving solely due to gravitational
radiation reaction.

\section{Rescaling the Disk to the Halo}\label{sec:RenormDisk}
A simple estimate of the gravitational wave background expected
from binary WDs in the halo can be obtained by assuming that the
WD binary population of the halo is similar in nature to that of
the disk.  One can estimate the halo GW background by rescaling
the disk binary WD population, based on three factors:
\begin{enumerate}
\item The ratio of the total number of halo WD MACHOs to the total
number of Galactic disk WDs, $N_{\rm halo}/N_{\rm disk}$.

\item The ratio of the average inverse distance squared of a halo
MACHO to the average inverse distance squared of a disk WD, $\langle
d^{-2} \rangle_{\rm halo}/ \langle d^{-2} \rangle_{\rm disk}$.

\item The ratio of the fraction of white dwarfs in binaries in the
halo to the fraction of white dwarfs in binaries in the disk,
$\alpha$.
\end{enumerate}

The number of WDs in the disk is computed by integrating over the
standard cylindrical exponential model,
\begin{equation}
    \rho = \rho_{0} \exp \left[ {-r \over r_{0}}\right]
    \exp \left[ {-|z| \over z_{0}}\right] ,
    \label{disk}
\end{equation}
where $r_{0} = 3.5$ kpc and $z_{o} = 90$ pc are the exponential
scale heights for the WD population in the disk (Hils, Bender \&
Webbink, 1990), and $\rho_{0} = 4.72 \times 10^{-2}\; {\rm
{pc}}^{-3}$ is the number density of white dwarfs at the center
of the galaxy (computed from the local density of white dwarfs in
the solar neighborhood, $\rho_{\odot} = 4.16 \times 10^{-3}\;
{\rm {pc}}^{-3}$ given in Knox, Hawkins \& Hambly, 1999).
Integrating the disk WD density using the distribution in Eq.\
(\ref{disk}) yields
\begin{equation}
    N_{\rm disk} = 6.5 \times 10^8
    \label{Ndisk}
\end{equation}
for the disk population.

We assume that the number of MACHOs in a $50$ kpc halo (to the
LMC), is $N_{\rm halo}^{50\; {\rm kpc}} = 2 \times 10^{11}$, the
halo-model-independent result obtained by the MACHO collaboration
(Alcock {\it et al.}, 2000).  For a larger halo, extending some
$300$ kpc (halfway to M31), this number can be scaled by assuming
that the spatial distribution of white dwarf MACHOs follows the
standard spherical flat rotational halo model given by
\begin{equation}
    \rho = {\hat \rho}\, {R^2+a^2 \over r^2 + a^2} \; \; ,
    \label{halo}
\end{equation}
where ${\hat \rho}$ is the local density of dark matter, $r$ is
the Galactocentric radius, $R = 8.5\; {\rm kpc}$ is the
Galactocentric radius of the Sun, and $a = 5.0 \; {\rm kpc}$ is
the halo core radius. Integrating Eq.\ (\ref{halo}) out to $50$
kpc and setting the number of MACHOs equal to $2 \times 10^{11}$,
one obtains ${\hat \rho} = 0.0094 \; \rm{M}_\odot\; {\rm
{pc}}^{-3}$. Using this value and integrating Eq.\ (\ref{halo})
out to 300 kpc gives
\begin{equation}
    N_{\rm halo}^{300\; {\rm kpc}} = 1.3 \times 10^{12} \; \; .
    \label{Nhalo}
\end{equation}

The average inverse distance squared between sources and the Sun,
$\langle d^{-2}\rangle $, may be found by integrating over the
source distributions given in Eqs.\ (\ref{halo}) and
(\ref{disk}).  Expressing the result as a distance, one finds that
for the disk,
\begin{equation}
    \langle d^{-2} \rangle^{-1/2} = 4.85 \;{\rm kpc} \; ,
    \label{d2disk}
\end{equation}
while
\begin{equation}
    \langle d^{-2} \rangle^{-1/2} = 15.7 \; {\rm kpc}
    \label{d2halo50}
\end{equation}
for a $50$ kpc halo, and
\begin{equation}
    \langle d^{-2} \rangle^{-1/2} = 39.5 \; {\rm kpc}
    \label{d2halo300}
\end{equation}
for a $300$ kpc halo.

Nothing is presently known about the ratio of the fraction of white
dwarfs in binaries in the halo to the fraction of white dwarfs in
binaries in the disk, $\alpha$, so we leave this as a free parameter
in our model, and examine the consequences of different values for
$\alpha$.

In rescaling the disk background GW spectrum, there are two
separate effects associated with the potentially larger number of
WD binaries in the halo.  First, the larger number tends to raise
the overall level of the halo background relative to that of the
disk. Second, since there are more halo binaries per unit
frequency interval, this pushes the point in the spectrum where
one first encounters open frequency bins (one or fewer Galactic
halo binaries per bin), to higher frequencies than in the disk.
Scaling the number of Galactic disk binaries by multiplying by
the ratio of the total number of WDs in the halo to the number in
the disk yields
\begin{equation}
    \log_{10} \left ( {dN \over df} \right )_{\rm halo} = - \left (
    {11 \over 3} \right ) \log_{10} (f) -2.35 + \log_{10} \left (
    \alpha N_{\rm halo} \over N_{\rm disk} \right ) \; \; .
    \label{dNdfhalo}
\end{equation}

The overall level of the GW backgrounds from Galactic halo WD
binaries and extragalactic halo binaries will scale according to
\begin{equation}
    h^{\rm halo}_{f} = K(\alpha) h^{\rm disk}_{f}\ ,
    \label{rescale}
\end{equation}
where the scaling factor $K(\alpha)$ is defined as
\begin{equation}
    K(\alpha) =\left[ \alpha
    {N_{\rm halo} \over N_{\rm disk}}
    {{\langle d^{-2} \rangle_{\rm halo}} \over
    {\langle d^{-2} \rangle_{\rm disk}}}\right]^{1/2}\ .
    \label{scale}
\end{equation}

If one assumes that the halo population of white dwarfs precisely
mimics the disk population, then the same fraction of WDs will be
in binaries in the halo as in the disk, and $\alpha = 1$.  In
this case the scaling factor is given by
\begin{equation}
      K(\alpha = 1) = \left\{
      \begin{array}{ll}
         5.42 & \mbox{for a 50 kpc halo}\\
         5.49 & \mbox{for a 300 kpc halo}
      \end{array} \; .
    \right.
  \label{zval}
\end{equation}

An estimate of the response of LISA to a background of GW from
halo WD binaries can now be obtained by rescaling the Galactic
and extragalactic disk spectral amplitudes [Eqs.\ (\ref{hfGdisk})
and (\ref{hfedisk})] using Eq.\ (\ref{rescale}) with Eq.\
(\ref{dNdfhalo}) in Eq.\ (\ref{effhf}).  The resulting signal
estimate for LISA is illustrated in Figure (\ref{ScaledSignal}),
along with the Bender-Hils estimate for disk binaries for
comparison.  The signal from the halo WD binaries is
substantially stronger than that from the disk binaries. At lower
frequencies, the predicted backgrounds for a 50 kpc and 300 kpc
halo are indistinguishable in the figure, owing to the similarity
in the values of $K$ in Eq.(\ref{zval}). The larger number of
halo binaries compared to the disk fills the frequency bins to a
substantially higher frequency before one encounters open bins,
where weaker signals such as the extragalactic background may be
observed. If only the disk background is present, then potential
extragalactic sources can be detected above a critical frequency
of about $2 \times 10^{-3}$ Hz, where frequency bins cease to be
cluttered with many Galactic binaries. With a 50 kpc halo, the
greater number of Galactic binaries increases this critical
frequency to about $1 \times 10^{-2}$ Hz, while for a larger 300
kpc halo, the critical frequency is further increased to about $2
\times 10^{-2}$ Hz. In the latter case, the frequency at which
bins begin to open up and allow weaker extragalactic signals to
be detected is roughly equal to where LISA's instrumental noise
curve begins to rise.

Of course, it is presently unknown whether the fraction of WDs in
binaries in the halo is comparable to that in the disk.  A
priori, it could be larger ($\alpha > 1$) or smaller ($\alpha <
1$).  One question which can be posed within the present simple
model is to ask what value of $\alpha$ will result in the halo GW
signal being similar in magnitude and shape to that of the disk.
This determines a minimum value for $\alpha$, above which the
halo signal will dominate over that of the disk binaries.
Reducing $\alpha$ in Eqs.(\ref{dNdfhalo}) and {\ref{rescale}) one
finds that the signal from a $50$ kpc halo will be dominant if
$\alpha > 10^{-2}$, while the signal strength from a $300$ kpc
halo would exceed that of the disk if $\alpha > 5 \times
10^{-3}$. This implies that even if the fraction of halo WDs in
binaries is as low as 1\% of the fraction of disk WDs in
binaries, the WD MACHO binaries will be ``bright'' enough to
stand out from the expected signal of the disk binaries.  While
the numbers here are highly sensitive to the specific details of
the halo binary MACHO population, they illustrate that the halo
will be a significant source of a low frequency confusion
background of gravitational waves unless the fraction of MACHO
WDs in binaries is orders of magnitude lower than in the disk.

\section{Discussion}\label{sec:Discussion}
While this simple scaled model is certainly not an accurate
representation of the Galactic halo binaries, it does illustrate
that the GW background from a halo population of white dwarf
binaries could easily dominate the signal in a space-based
interferometer such as LISA. Further, this result, together with
other studies that have considered the GW signal from MACHOs if
they were identified with primordial black holes (Nakamura {\it
et al.} 1997; Hiscock, 1998; Ioka {\it et al.}, 1999; Ioka {\it
et al.}, 2000), demonstrates that whatever the nature of MACHOs,
if even a small fraction of them are in binary systems, then they
will create a strong confusion noise background which could
saturate the frequency range in which LISA is most sensitive ($ 2
\times 10^{-3} \; {\rm Hz} < f < 2 \times 10^{-2} \; {\rm Hz}$).

Some may consider such a prospect discouraging, as the combined
signal from abundant halo binaries could mask other weak signals
and make them undetectable.  This has previously been a serious
concern with respect to the disk binaries--hence the name
``confusion noise'' for a signal that actually describes the
short-period binary population of the Galaxy.  There has been
hope that the confusion noise from Galactic disk binaries could
be extracted from the general stochastic (e.g., cosmological)
background by utilizing the anisotropic nature of the disk signal
(Giampieri \& Polnarev, 1997).  If there is a substantial signal
associated with halo binaries, however, then this scheme will not
work.  The solar position is sufficiently near the center of a
spherical Galactic halo that it would seem difficult or
impossible to be able to subtract out the halo confusion noise
signal based on its very small anisotropy.

On the other hand, even our simple analysis shows that the GW
signal from the halo binaries could be a powerful tool to
determine the properties of the MACHO binary population, as well
as general properties of the halo itself, such as its size.

\acknowledgments The authors wish to thank P.\ Bender and R.\ Hellings
for helpful discussions.  This work was supported in part by NSF Grant
No.\ PHY-9734834 and NASA Cooperative Agreement NCC5-410.

\newpage

\begin{figure}
\caption{The expected response of LISA to the gravitational wave
signal from a halo population of white dwarf binaries is shown.
The upper solid curve represents a 50 kpc halo; the dashed curve
represents a 300 kpc halo.  The lower curve is the detailed
spectrum predicted by Bender and Hils for disk binaries. The
expected sensitivity of the LISA interferometer (S:N = 1) is also
shown.} \label{ScaledSignal}
\end{figure}

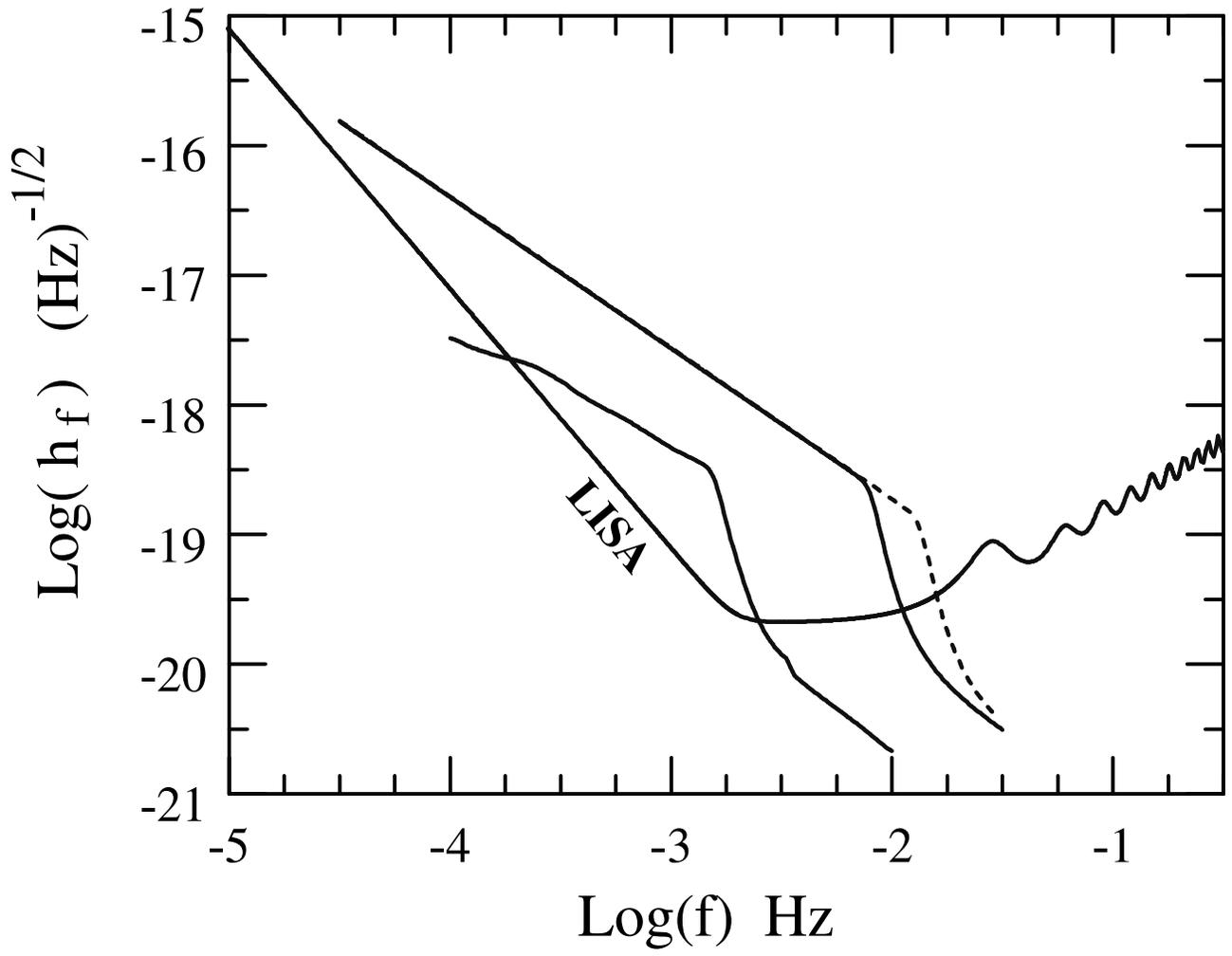
\end{document}